\documentclass[prd,twocolumn,tightenlines,showpacs,nofootinbib,superscriptaddress]{revtex4-1} 

\usepackage{amsfonts}
\usepackage{dcolumn}
\usepackage{bm}
\usepackage{amsmath}
\usepackage{nicefrac}
\usepackage{amsthm}
\usepackage{amssymb}
\usepackage{graphicx}
\usepackage{color}

\newcommand{\appropto}{\mathrel{\vcenter{
  \offinterlineskip\halign{\hfil$##$\cr
    \propto\cr\noalign{\kern2pt}\sim\cr\noalign{\kern-2pt}}}}}

\begin{document}

\title{Reply to comment on ``Searching for Topological Defect Dark Matter via Nongravitational Signatures''}

\date{\today}
\author{Y.~V.~Stadnik} 
\affiliation{School of Physics, University of New South Wales, Sydney 2052, Australia}
\author{V.~V.~Flambaum} 
\affiliation{School of Physics, University of New South Wales, Sydney 2052, Australia}
\affiliation{Mainz Institute for Theoretical Physics, Johannes Gutenberg University Mainz, D 55122 Mainz, Germany}

\begin{abstract}
In the comment of Avelino, Sousa and Lobo [arXiv:1506.06028], it is argued, by comparing the kinetic energy of a topological defect with the overall energy of a pulsar, that the origin of the pulsar glitch phenomenon due to the passage of networks of topological defects through pulsars is faced with serious difficulties. 
Here, we point out that topological defects may trigger pulsar glitches within traditional scenarios, such as vortex unpinning. 
If the energy transfer from a topological defect exceeds the activation energy for a single pinned vortex, this may lead to an avalanche of unpinning of vortices and consequently a pulsar glitch, and therefore the source of angular momentum and energy required for a glitch event is provided by the pulsar itself.
Indeed, the activation energy for such a process can be very small (essentially zero compared with the observed increase in the pulsar's rotational kinetic energy at the onset of a glitch). 
The unpinning of a vortex by a topological defect may occur through the passage of the defect into the core of the pulsar.
\end{abstract}


\maketitle 

A pulsar glitch is an abrupt increase in the rotational frequency of a pulsar, typically of magnitude $\delta \Omega / \Omega \sim 10^{-11} - 10^{-5}$ and over a temporal interval of at least $\sim 30$ s, with a comparatively slow recovery period (not necessarily to the original state of rotation) ranging from $T_{\textrm{d}} \sim 1$ day $-$ 3 yr, see, e.g., Refs.~\cite{Dodson2007,Espinoza2011,ATNF}. Various models attempting to explain pulsar glitches have been proposed, but the exact origin of this phenomenon is still disputed, see, e.g., the review in Ref.~\cite{Haskell2015}.

In Ref.~\cite{Avelino2015comment}, it is argued, by comparing the kinetic energy of a topological defect (TD) with the overall energy of a pulsar, that the origin of the pulsar glitch phenomenon due to the passage of networks of TDs through pulsars is faced with serious difficulties, at least in the case of a repulsive interaction between a TD and a pulsar when the TD cannot immediately penetrate the pulsar (in the case of an attractive interaction, a TD has potential energy in addition to kinetic energy, and kinetic energy arguments are not valid).
Here, we point out that the mechanism for a pulsar glitch may be completely different to that considered in Ref.~\cite{Avelino2015comment}. 
In particular, TDs may trigger pulsar glitches within traditional scenarios, such as the unpinning of vortices (which carry angular momentum). 
If the energy transfer from a TD exceeds the activation energy for a single pinned vortex, this may lead to an avalanche of unpinning of vortices and consequently a pulsar glitch due to the transfer of large amounts of angular momentum from the superfluid core to the crust; therefore, the source of angular momentum and energy required for a glitch event is provided by the pulsar itself.
Indeed, the activation energy for such a process can be very small, with pinning energies $E_{\textrm{pin}} \sim 1$ MeV for a single vortex being quite typical (see e.g.~\cite{Pines1991book,Blaschke2001LNP} and references therein). Such pinning energies are essentially zero when compared with the rotational kinetic energy of a pulsar:~$E_{\textrm{rot}} \sim 10 ^{56} - 10^{64}$ eV (the increase in the pulsar's rotational kinetic energy at the onset of a glitch is $\delta E_{\textrm{rot}} = 2 E_{\textrm{rot}}  \delta \Omega / \Omega$, with $\delta \Omega / \Omega \sim 10^{-11} - 10^{-5}$).
The unpinning of a vortex by a TD may occur through the passage of the TD into the core of the pulsar if the TD's interaction with the pulsar is attractive. In the case of a repulsive interaction, there may be a possibility of delayed energy transfer, which may occur due to the collision of a pulsar with an extended TD such as a domain wall. A moving domain wall will be stretched by the pulsar if it cannot immediately pass through due to the repulsion. If the elastic force due to the stretching of the domain wall exceeds the repulsive force from the pulsar, the wall penetrates inside the pulsar and may cause vortex unpinning. 
To realise this possibility in nature, there should be a high mass-density domain wall.
Note that the acceleration of the pulsar by the domain wall also leads to a change in the pulsar's rotational frequency in the reference frame of an external observer. 


Further progress into this problem requires detailed numerical investigations. TDs are assumed to be identical (similar to elementary particles). Therefore, we should calculate the responses of different pulsars to the same TD and compare with observational data. The findings of this investigation will be presented in a future publication. 
Finally, we note that, if pulsar glitches are caused by TDs, then this may be independently corroborated through correlated astrophysical measurements involving a network of pulsars, or better yet binary pulsar systems, further to our original proposals in Ref.~\cite{Stadnik2014defects}, as well as recently proposed terrestrial experiments to search for TDs using a global network of atomic clocks \cite{Derevianko2014} and laser interferometers \cite{Stadnik2015laser,Stadnik2015laserB}.

\textbf{Acknowledgements} --- This work was supported by the Australian Research Council.



\end{document}